\documentclass[manuscript,screen,sigconf]{acmart}
\AtBeginDocument{%
  \providecommand\BibTeX{{%
    \normalfont B\kern-0.5em{\scshape i\kern-0.25em b}\kern-0.8em\TeX}}}

\copyrightyear{2023} 
\acmYear{2023} 
\setcopyright{acmlicensed}\acmConference[SC-W 2023]{Workshops of The International Conference on High Performance Computing, Network, Storage, and Analysis}{November 12--17, 2023}{Denver, CO, USA}
\acmBooktitle{Workshops of The International Conference on High Performance Computing, Network, Storage, and Analysis (SC-W 2023), November 12--17, 2023, Denver, CO, USA}
\acmPrice{15.00}
\acmDOI{10.1145/3624062.3624195}
\acmISBN{979-8-4007-0785-8/23/11}

\begin{document}

%%
%% The "title" command has an optional parameter,
%% allowing the author to define a "short title" to be used in page headers.
\title{Evaluating the performance portability of SYCL across CPUs and GPUs on bandwidth-bound applications}

%%
%% The "author" command and its associated commands are used to define
%% the authors and their affiliations.
%% Of note is the shared affiliation of the first two authors, and the
%% "authornote" and "authornotemark" commands
%% used to denote shared contribution to the research.
\author{Istv\'an Z Reguly}
\email{reguly.istvan@itk.ppke.hu}
\orcid{0000-0002-4385-4204}
\affiliation{%
  \institution{P\'azm\'any P\'eter Catholic University, Faculty of Information Technology and Bionics}
  \streetaddress{Pr\'ater utca 50/a}
  \city{Budapest}
  %$\state{Ohio}
  \country{Hungary}
  \postcode{1083}
}
\renewcommand{\shortauthors}{Istvan Reguly}

\begin{abstract}
  In this paper, we evaluate the portability of the SYCL programming model on some of the latest CPUs and GPUs from a wide range of vendors, utilizing the two main compilers: DPC++ and hipSYCL/OpenSYCL. Both compilers currently support GPUs from all three major vendors; we evaluate performance on the Intel(R) Data Center GPU Max 1100, the NVIDIA A100 GPU, and the AMD MI250X GPU. Support on CPUs currently is less established, with DPC++ only supporting x86 CPUs through OpenCL, however, \\ OpenSYCL does have an OpenMP backend capable of targeting all modern CPUs; we benchmark the Intel Xeon Platinum 8360Y Processor (Ice Lake), the AMD EPYC 9V33X (Genoa-X), and the Ampere Altra platforms. We study a range of primarily bandwidth-bound applications implemented using the OPS and OP2 DSLs, evaluate different formulations in SYCL, and contrast their performance to “native” programming approaches where available (CUDA/HIP/OpenMP). On GPU architectures SCYL on average even slightly outperforms native approaches, while on CPUs it falls behind - highlighting a continued need for improving CPU performance. While SYCL does not solve all the challenges of performance portability (e.g. needing different algorithms on different hardware), it does provide a single programming model and ecosystem to target most current HPC architectures productively.
  %to make SYCL a truly performance portable programming model.
\end{abstract}



%%
%% Keywords. The author(s) should pick words that accurately describe
%% the work being presented. Separate the keywords with commas.

% \received{20 February 2007}
% \received[revised]{12 March 2009}
% \received[accepted]{5 June 2009}

%%
%% This command processes the author and affiliation and title
%% information and builds the first part of the formatted document.
\maketitle

\section{Introduction}

Performance portability of scientific codes has been an increasing challenge ever since the emergence of accelerators. There has been an explosion of programming approaches that target some of the existing hardware architectures - developers of high performance computing (HPC) codes on one hand want to exploit the performance offered by the latest and greatest architectures, but on the other hand cannot afford to keep refactoring their codes, or even keep multiple implementations around.

Over time, it has become clear that the level of abstraction of programming models (lower level models expose a wider range of capabilities) is proportional to their portability; e.g. CUDA allows access to the full range of features on NVIDIA GPUs, but is not portable to any other vendors. Approaches at higher levels of abstraction, such as OpenMP do enable a better degree of portability, though still not trivially, as it uses different pragmas and clauses for CPUs and offload targets. KOKKOS \cite{kokkos} represents an even higher level of abstraction, and is portable across most shared-memory parallel architectures. DSLs, such as OP2 and OPS \cite{op2,ops} further increase the level of abstraction, narrowing it to two of the computational dwarfs \cite{asanovic2006landscape}, but allow portability not just shared memory, but distributed memory systems too.

At the lower levels of abstraction (SIMT - Single Instruction Multiple Threads) there was an early effort in OpenCL \cite{munshi2009opencl} to deliver portability, but due to less than enthusiastic support from hardware vendors and the HPC community, it never took a significant share of used programming models in the HPC space. The SYCL and SYCL 2020 standards adopted a modern C++ interface to make the SIMT model more easily available. Intel's OneAPI effort in particular allows a single source code to be portable across most modern HPC platforms, including support for native hardware libraries (such as BLAS) with a single interface - allowing much easier access to a large existing software ecosystem.

\begin{table*}[]
\begin{tabular}{|l|l|l|l|l|l|}
\hline
\begin{tabular}[c]{@{}l@{}}AMD  \\ MI250X\\ (1 GCD)\end{tabular} & \begin{tabular}[c]{@{}l@{}}NVIDIA\\ A100 40GB\end{tabular} & \begin{tabular}[c]{@{}l@{}}Intel Data Center\\ GPU Max 1100\end{tabular} & \begin{tabular}[c]{@{}l@{}}Intel Xeon\\ Platinum 8360Y\\ dual-socket\end{tabular} & \begin{tabular}[c]{@{}l@{}}AMD EPYC\\ 9V33X\\ dual-socket\end{tabular} & \begin{tabular}[c]{@{}l@{}}Ampere\\ Altra\\ \\ single socket\end{tabular} \\ \hline
1290 GB/s                                                        & 1310 GB/s                                                  & 803 GB/s                                                                 & 296 GB/s                                                                          & 561 GB/s                                                               & 167 GB/s                                                                  \\ \hline
\end{tabular}
\vspace{-8pt}
\caption{Achieved bandwidth on STREAM Triad, measured with BabelStream \label{tab:bw}\vspace{-20pt}}.
\end{table*}

As part of OneAPI, Intel develops the OneAPI DPC++/C++ compiler (to which we will refer to as the DPC++ compiler later on), a version of which is available open source \cite{dpcpp}, currently supports Intel/NVIDIA/AMD GPU architectures by directly compiling to their respective intermediate representations or to SPIR-V\cite{spirv}. CPU architectures are supported by targeting OpenCL drivers capable of accepting SPIR-V - currently the only well-established drivers are from Intel, targeting x86 CPUs. The OpenSYCL \cite{hipsycl} project similarly compiles code to device-specific IRs in LLVM for GPUs, but is also capable of utilizing OpenMP to target any CPUs, including ARM.

There have been numerous papers exploring porting CUDA or other codes to SYCL \cite{tsai2021porting,christgau2020porting}, and some that  study portability across different platforms \cite{jin2022understanding, goli2020towards} for a particular application. Our work is novel in that it takes a range of different bandwidth-bound applications (7 in total), and evaluates performance portability across the widest range of hardware architectures yet, including the latest NVIDIA A100, AMD MI250X, and Intel Data Center GPU Max 1100 GPUs, as well as some of the latest avilable CPUs from Intel (Ice Lake), AMD (Genoa-X), and ARM (Ampere Altra). We study the effects of different formalizations of SYCL code on performance across different hardware and compilers, and contrast them to ``native'' programming approaches.

The rest of the paper is structured as follows: Section 2 introduces the hardware platforms studied, Section 3 presents the applications and their parallel implementations that we evaluate, Section 4 conducts a detailed analysis of results, comparing applications, parallelizations, and compilers. Finally, Section 5 draws conclusions.

\section{Test hardware}

To conduct this study, the following hardware platforms and software configurations were evaluated:

\begin{enumerate}
    \item Intel Xeon Platinum 8360Y Processor, available in the Baskerville cluster at the University of Birmingham. Two sockets, each with 36 cores, Hyperthreading on. 512 GB DDR4 RAM. Clock frequencies between 2.4 GHz (base frequency) - 2.8 GHz (all-core turbo), giving a theoretical 11-13 FP32 TFLOPS/s. Software: RHEL 8.5, Intel OneAPI Base and HPC toolkits, 2023.1 (including Intel MPI). Benchmarked July 11, 2023.
    \item AMD EPYC 9V33X with 3D V-Cache Technology, available as an Azure HB176rs v4 virtual machine. Two sockets, each with 88 available cores, Hyperthreading off. 2x2 NUMA regions, with 704 GB DDR5 RAM. Clock frequencies between 2.4 GHz (base frequency) - 3.7 GHz (turbo), giving a theoretical 9.22-14.22 FP32 TFLOPS/s. Software: Ubuntu 22.04, GCC 12.3 and AMD Optimizing C/C++ Compiler 4.0. Benchmarked Aug 4, 2023.
    \item Ampere Altra, available as an Azure D64ps v5 virtual machine. One socket with 64 available cores, Hyperthreading off. Single NUMA region, with 208 GB DDR4 RAM. Clock frequency at 3.0 GHz, giving a theoretical 3 FP32 TFLOPS/s. Software: Ubuntu 22.04, GCC 12.3 and ARM Compiler 23.04.1 Compiler Benchmarked Aug 5, 2023.
    \item AMD MI250X GPU, available on the LUMI (Cray EX) supercomputer. Only a single GCD was benchmarked. 110 compute units, 7040 streaming processors, running at up to 1700 MHz, giving 23.95 peak FP32 TFLOPS/s. Software: RoCM 5.4.2, Cray Programming Environment 22.12. Benchmarked Apr 15, 2023. %L2 16MB
    \item NVIDIA A100 40GB PCI-e GPU, 108 streaming processors, 6912 CUDA cores, running at up to 1410 MHz, giving 19.49 peak FP32 TFLOPS/s. Software: Centos 8 Stream, CUDA 11.6, Benchmarked Jun 10, 2023. %L2 40MB
    \item Intel Data Center GPU Max 1100, available in the Intel Developer Cloud. 56 $X^e$ cores, running at up to 1550 MHz. Software: Ubuntu 22.04, Intel OneAPI Base and HPC toolkits, 2023.1 (including Intel MPI). Benchmarked Aug 5, 2023. %L2 204 MB
\end{enumerate}

Since the applications studies are primarily bound by memory bandwidth, the achievable peak bandwidth on these platforms is of key interest. To measure this, we used BabelStream \cite{babelstream}, compiled with the native parallelizations and compilers, and report them in Table \ref{tab:bw} - we will use these figures later to compute achieved fraction of peak, and refer it to as \textit{achieved architectural efficiency}.

% \begin{table}[]
% \begin{tabular}{|l|l|l|}
% \hline
% AMD MI250X     & NVIDIA A100       & Intel GPU Max 1100 \\ \hline
% 1290 GB/s  & 1310 GB/s  & 803 GB/s \\ \hline
% Intel Xeon 8360Y & AMD EPYC 9V33X & Ampere Altra    \\ \hline
% 296 GB/s   & 561 GB/s   & 167 GB/s \\ \hline
% \end{tabular}
% \caption{Achieved bandwidth on STREAM Triad, measured with BabelStream \label{tab:bw}}.
% \end{table}

\section{Applications and parallelizations}

The benchmarked applications include a range of primarily bandwidth-limited applications, with varying levels of computational intensity and complexity. Structured mesh applications are structurally simpler, whereas the MG-CFD unstructured mesh requires more complex addressing and the resolution of race conditions. The codes and their key properties as are follows:

\label{sec/apps}
\begin{enumerate}
\item CloverLeaf 2D/3D \cite{cloverleaf} – structured-mesh Eulerian hydrodynamics simulation, representative of nuclear security codes. Mostly bandwidth-bound, with some operations on faces/edges that may be latency bound. Double precision, $7680^2$(2D), $408^3$(3D) problem size, 50 iterations.
\item OpenSBLI SA \& SN \cite{opensbli} – structured mesh finite difference Navier-Stokes solver for capturing shock-boundary layer interactions. Production code with 2 varians – Store All (SA), which is bandwidth-bound, and Store None (SN), which recomputes derivatives on the fly, reducing data movement pressure, but still mostly bandwidth bound. Double precision, $320^3$ problem size, 20 time iterations.
\item RTM - proxy code for a Reverse Time Migration application's forward pass, uses a complex 8th order finite difference stencil. Sensitive to cache locality and vectorization, has large communications volume over MPI. Single precision, $320^3$ problem size, 10 time iterations.
\item Acoustic – structured-mesh high-order (8th) finite difference acoustic wave propagation solver. Bandwidth and cache locality bound, with large communications volume over MPI. Single precision, $1000^3$ problem size, 30 time iterations.
\item MG-CFD \cite{mgcfd} – unstructured mesh finite volume Euler equations solver with multigrid – proxy for Rolls-Royce’s CFD simulator Hydra. Bound by latencies and indirect memory accesses. Double precision, NASA Rotor37 case with 8 million vertices, 25 iterations.
% \item miniBUDE \cite{miniBUDE} – proxy molecular docking code, representative of BUDE. Compute and latency bound. bm1 testcase, 30 iterations.
%$\item Volna \cite{volna} – unstructured mesh finite volume Nonlinear Shallow Water Equations solver. Also sensitive to indirect memory accesses as MG-CFD, but less so. Indian ocean case with 30 million vertices, 200 time iterations.
%\item miniWeather \cite{miniweather} – structured mesh proxy code implementing basic dynamics seen in atmospheric weather and climate simulations. Bandwidth bound. $4000x2000$ problem size, simulation time 1.0.
\end{enumerate}

All codes are implemented in the OPS (structured mesh) and OP2 (unstructured mesh) Domain Specific Languages \cite{ops,op2}, which allows us to generate different parallelizations based on the high-level description of the computations. As presented in earlier work, OP2 and OPS have been shown to deliver performance that closely matches hand-written and tuned implementations \cite{mudalige2015performance, kirk2017achieving, ops}. Both DSLs support the MPI and MPI+X execution models, where X can be OpenMP, CUDA/HIP, SYCL, and more.

OPS follows a standard Cartesian decomposition for structured mesh stencil codes, with shared memory parallelizations of 2D/3D loop nests. For SYCL specifically, we can generate two variants: a ``flat'' parallel implementation, where we pass a single \texttt{cl::sycl::range} to \texttt{parallel\_for}, as well as an ``nd\_range'' version, where workgroup shape is specified by passing a \texttt{cl::sycl::nd\_range} to \texttt{parallel\_for}. The key difference between the two approaches is that the ``flat'' version leaves the exact choice of workgroup shape for each individual kernel up to the runtime. The ``nd\_range'' version in contrast requires the programmer to specify the shape, and whereas this can be done for each an every kernel separately in principle, in our tests we only tune for the best performing shape for the entire application.

\begin{figure}[h]
    \centering
    \includegraphics[width=\columnwidth]{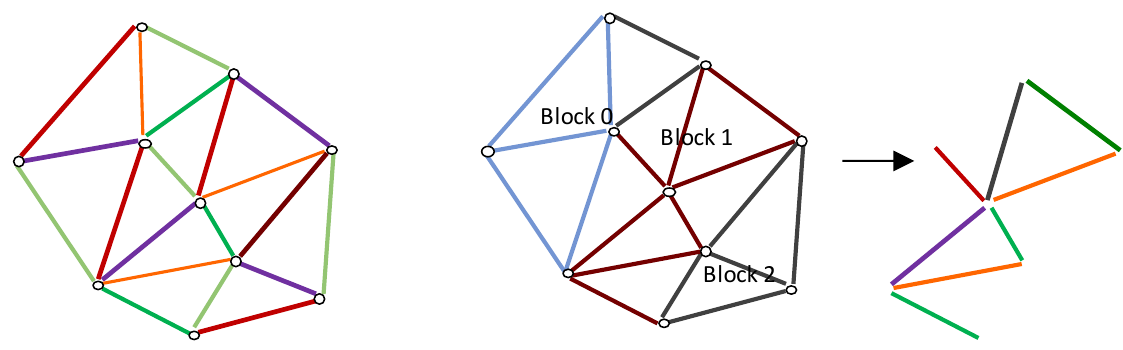} 
    \caption{Colored execution strategies for unstructured mesh computations - global (left) and hierarchical (middle and right)}
    \label{fig:coloring}
\end{figure}

Unstructured mesh computations on the other hand introduce data-driven dependencies: there are mapping tables describing connectivity between different sets (e.g. edges to vertices). In distributed systems, the problem is decomposed using a graph partitioner such as PT-Scotch \cite{chevalier2008pt}, and uses a standard owner-compute approach \cite{op2}. In shared memory parallel environment there are race conflicts to avoid: for example, when executing two edges that indirectly increment data on the same vertex. OP2 supports three execution strategies to resolve these conflicts, as illustrated on Figure \ref{fig:coloring}: (1) global coloring, where edges are colored so no two edges share the same vertex - this approach is simple, but by construction has very poor data reuse. (2) hierarchical coloring, where edges are broken up into groups (blocks), blocks are colored so no two blocks of the same color share a vertex, and subsequently edges within each block are colored as well - in hierarchical parallel environments (GPU threads and thread blocks) this allows for data reuse within blocks. (3) the use of atomics, where available and performant - this approach allows for good data locality (given a good mesh ordering) but may be limited by the throughput of atomic operations.

\section{Performance results}

\subsection{Structured mesh applications on GPU architectures}
\begin{figure*}[h]
    \centering
    \includegraphics[width=0.9\textwidth]{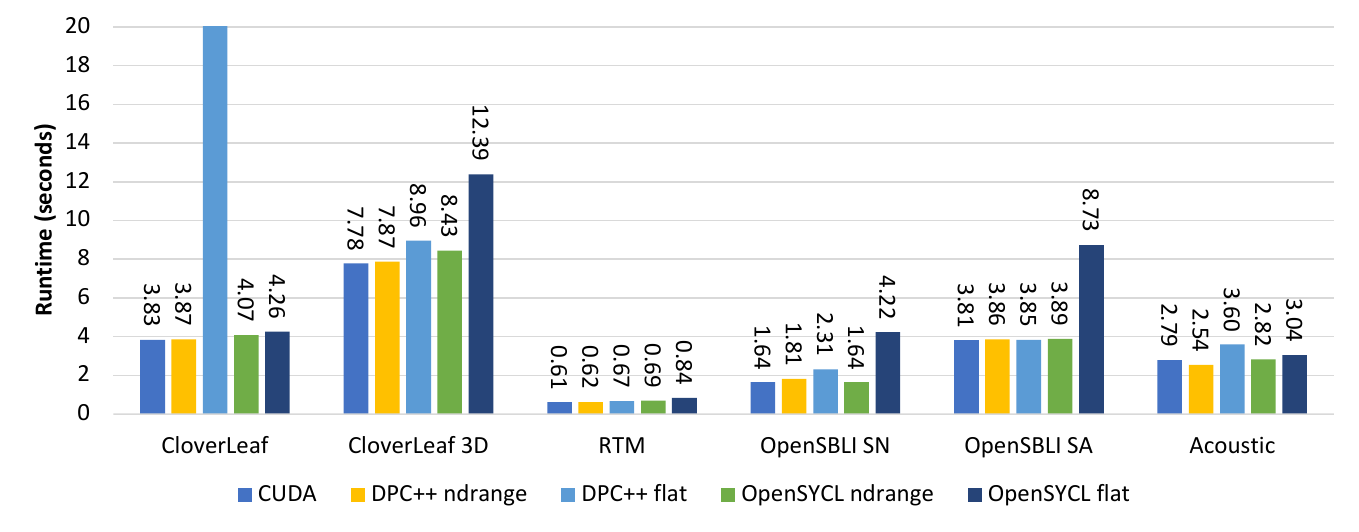} 
    \caption{Runtime of structured mesh applications on the NVIDIA A100 GPU with different compilers and parallel implementations}
    \label{fig:a100}
\end{figure*}

\begin{figure*}[h]
    \centering
    \includegraphics[width=0.9\textwidth]{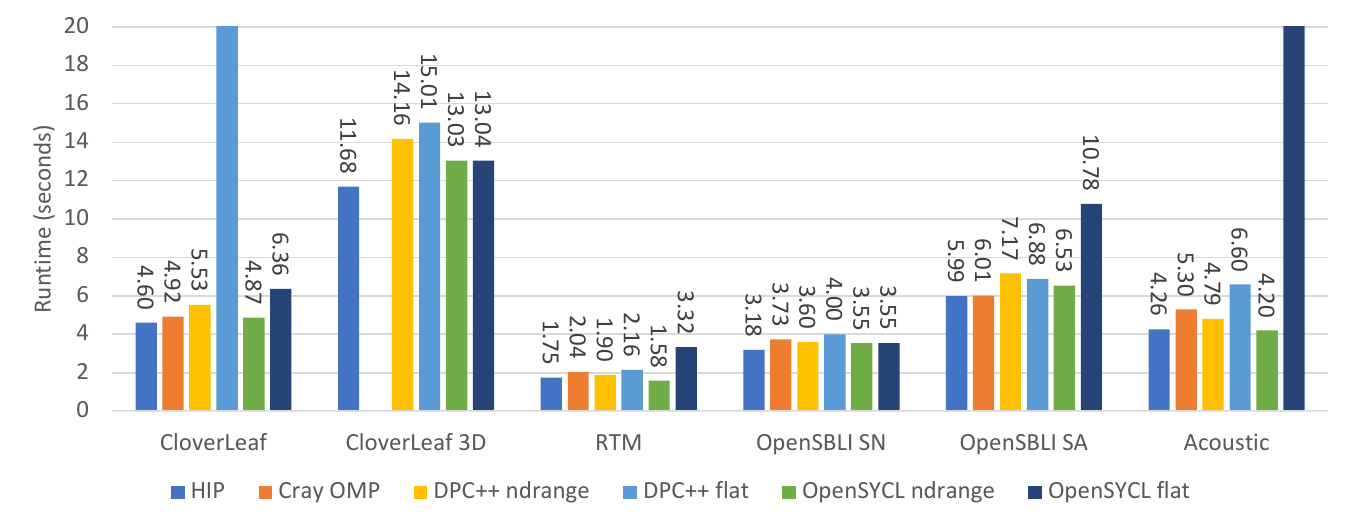} 
    \caption{Runtime of structured mesh applications on the AMD MI250X GPU (1 GCD) with different compilers and parallel implementations}
    \label{fig:mi250}
\end{figure*}

\begin{figure*}[h]
    \centering
    \includegraphics[width=0.9\textwidth]{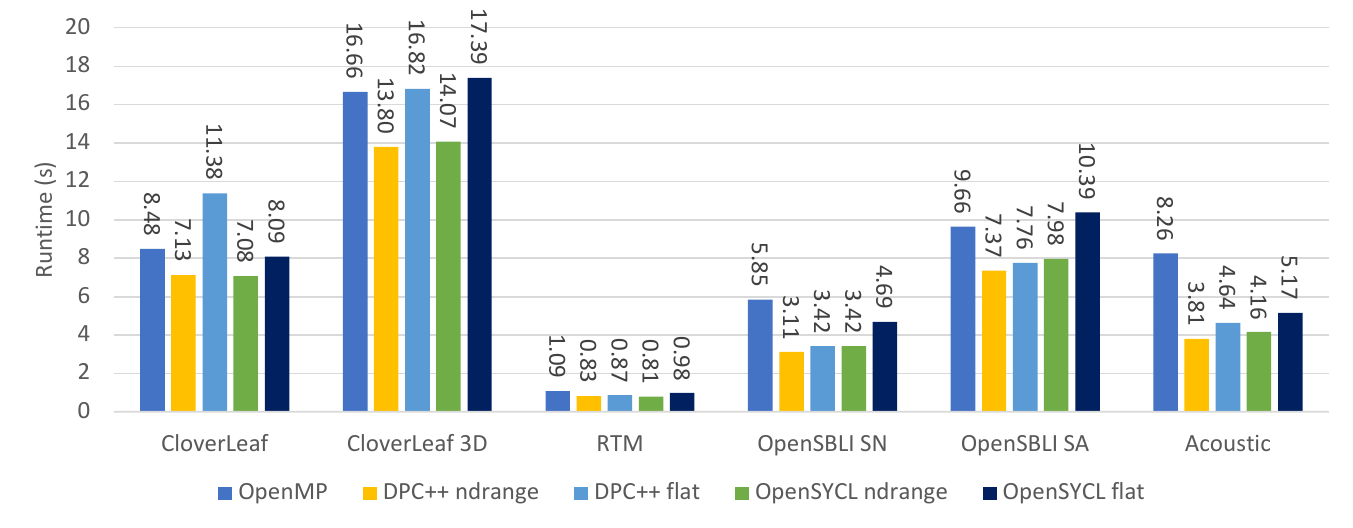} 
    \caption{Runtime of structured mesh applications on the Intel GPU Max 1100 with different compilers and parallel implementations}
    \label{fig:max1100}
\end{figure*}

The structurally simpler problems fall in the category of structured mesh stencil computations - although there is still significant diversity between applications in terms of computational intensity and cost of handling boundary conditions. Overall runtime results are shown in Figures \ref{fig:a100}, \ref{fig:mi250}, and \ref{fig:max1100} for the different applications and parallelizations.

CloverLeaf 2D and 3D have low computational intensity, very simple stencils, and a large fraction of loops operating on the boundary of the domain. The NVIDIA A100 achieves up to 92\% and 82\% architectural efficiency respectively - with 1.5\% and 7.8\% of time spent in boundary update kernels. While the native CUDA does perform best, the SYCL nd\_range versions with both compilers are within 10\%. The DPC++ runtime chooses very poor workgroup sizes for a few kernels, making the 2D version with the flat formulation perform very poorly - similarly the OpenSYCL version chooses suboptimal workgroup sizes in 3D, resulting in an almost 50\% slowdown. The RTM and Acoustic applications both achieve up to 48\% architectural efficiency - here SYCL compiled with DPC++ is highly competitive, outperforming CUDA on Acoustic by 10\%. On OpenSBLI, the more compute-intensive formulation StoreNone (SN) achieves 74\% efficiency, whereas the StoreAll (SA) variant achieves 92\% - there is significantly less variation for the less complex SA version in performance between different compilers and variations compared to the SN version, with only OpenSYCL + flat underperforming due to poor workgroup size choice. On average, the DPC++ compiler (with nd\_range) is only 1.2\% slower than native CUDA, and OpenSYCL with nd\_range is 5.3\% slower.
%halos mi250 a100 max1100
%c2d 2.57 1.5 0.9
%c3d 11.08 7.8 4.8

Performance on the AMD MI250X structurally paints a similar picture, with SYCL flat formulations performing poorly in similar combinations as on the A100. However, in contrast to the A100, the achieved architectural efficiency is consistently lower, with CloverLeaf 2D/3D at 78\% and 56\% - computations on the boundary take significantly longer (2.6\% and 11.1\% respectively), due to higher kernel launch latencies.  Similarly, RTM and Acoustic achieve 19\% and 30\% respectively, OpenSBLI SN achieves 39\% and SA 59\% efficiency. Both the DPC++ compiler and OpenSYCL offer competitive performance compared to HIP, particularly the optimized nd\_range variants. Notably on the RTM and Acoustic applications OpenSYCL performs better than the native HIP by 1.5\% and 11\% respectively. Figure \ref{fig:mi250} also shows runtimes achieved with OpenMP offload, compiled with the Cray compilers, showing competitive performance (though failing on CloverLeaf 3D). On average, the DPC++ compiler (with nd\_range) is 15.9\% slower than native HIP, and OpenSYCL with nd\_range is 4.5\% slower. In comparison to OpenMP offload code compiled with Cray, DPC++ is 2.3\% slower, but OpenSYCL is 9.1\% faster.

On the Intel Data Center GPU Max 1100, the recommended programming approach is SYCL, but we also compare to OpenMP offload variants (marked as ``native'' on plots), compiled with the OneAPI C/C++ compilers. Overall, efficiency on the Max 1100 is close to the NVIDIA A100 GPU - with CloverLeaf 2D/3D achieving 82\% and 72\% respectively, spending the least amount of time in boundary computations (0.9 and 4.8\%). The RTM and Acoustic applications achieve the highest fraction of peak across all GPUs (and even CPUs, except for Acoustic on the Genoa-X) at 59\% and 53\% respectively. The OpenMP offload variant and the SYCL flat formulations perform consistently worse than the tuned nd\_range  variants - and this difference is larger compared to the MI250X and A100 GPUs (ignoring the extremely poorly performing outliers). This is mainly because the Max 1100 is more sensitive to the right choice of workgroup shape, which based on our profiling comes down to L1/L2 cache hit rates improving significantly - this is supported by the fact that the Max 1100 has the largest L2 cache (at 208 MB), whereas the A100 only has 40 MB, and the MI250X 16 MB. On average, the DPC++ compiler with nd\_range is 30.2\% faster than OpenMP offload, and OpenSYCL is 27.6\% faster.

When looking at the consistency of achieved performance by taking the standard deviation of efficiencies of the best variant on each application, the Max 1100 has an average of 68\% (third-best after Genoa-X's 79\% and the A100's 73\%), but the lowest standard deviation, at 11.6\% (with the Xeon following at 11.8\%, and the rest above 17\%).

\subsection{Structured mesh applications on CPU architectures}

\begin{figure*}[h]
    \centering
    \includegraphics[width=0.9\textwidth]{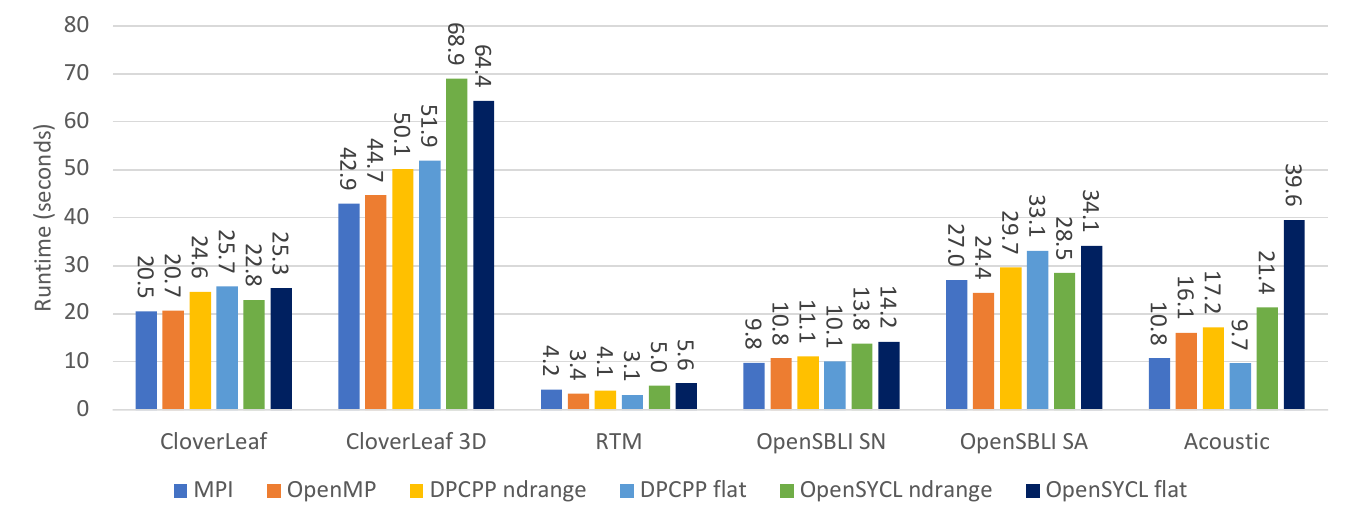} 
    \caption{Runtime of structured mesh applications on the Intel Xeon Platinum 8360Y Processor with different compilers and parallel implementations}
    \label{fig:xeon}
\end{figure*}

\begin{figure*}[h]
    \centering
    \includegraphics[width=0.9\textwidth]{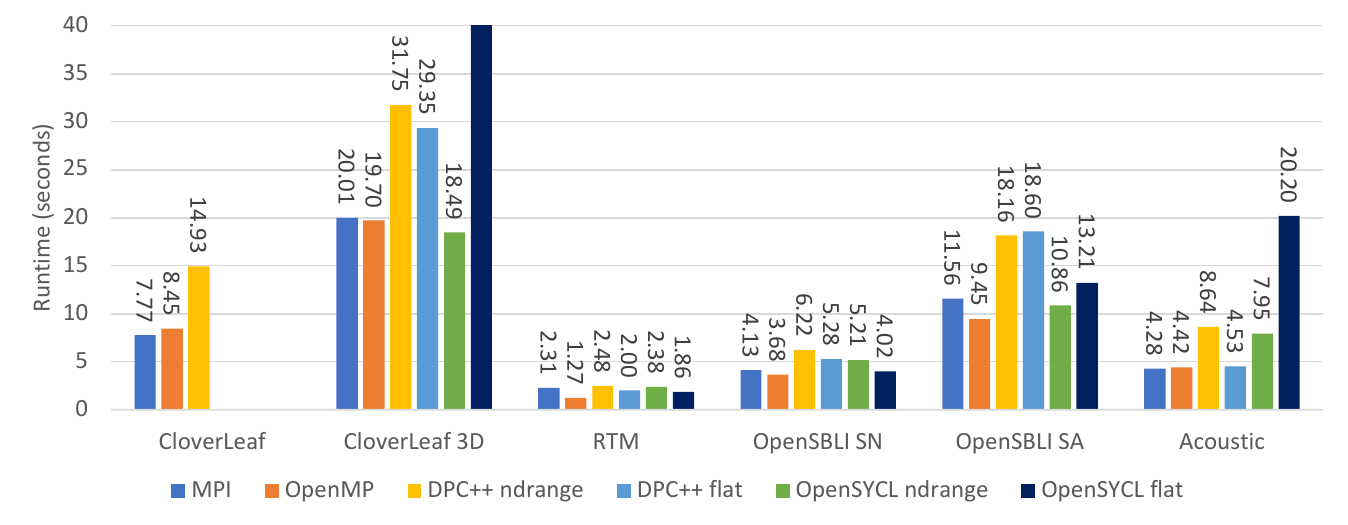} 
    \caption{Runtime of structured mesh applications on the AMD EPYC 9V33X Processor with different compilers and parallel implementations}
    \label{fig:genoax}
\end{figure*}

\begin{figure*}[h]
    \centering
    \includegraphics[width=0.9\textwidth]{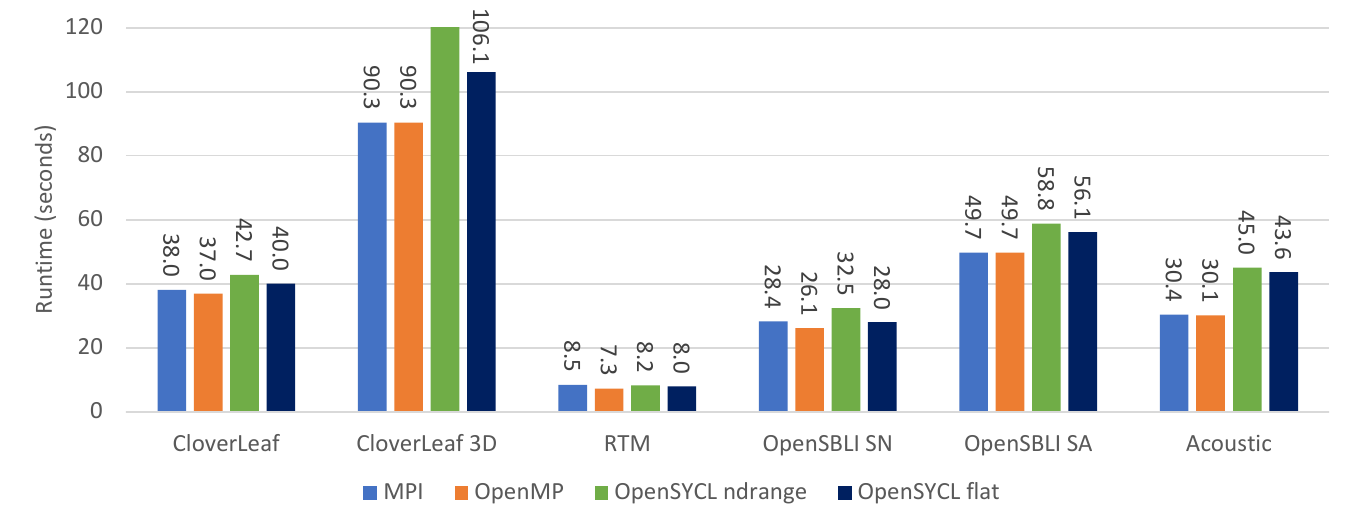} 
    \caption{Runtime of structured mesh applications on the Ampere Altra Processor with different compilers and parallel implementations}
    \label{fig:altra}
\end{figure*}

On CPU architectures the pure MPI and the hybrid MPI+OpenMP programming approaches are still the most prevalent in the high performance computing community, with OpenMP often utilized through a portability layer such as KOKKOS \cite{kokkos}. Therefore, as a baseline, we evaluate both pure MPI and MPI+OpenMP as implemented by OPS (forcing vectorization on the innermost loop, and collapsing the outer loops with OpenMP's \texttt{collapse} clause).

On the Intel Xeon Platinum 8360Y processor, we use the OneAPI C/C++ compilers (named icx/icpx) and Intel MPI (both from the 2023.1 toolkit) to compile MPI, OpenMP and SYCL (denoted with DPC++), and OpenSYCL compiled on top of LLVM/Clang 14.0, compiling applications with the \texttt{--opensycl-targets=omp.accelerated} switch, targeting LLVM's libomp. As on GPUs, the overhead of boundary condition handling is relevant here, because DPC++ has to go through the OpenCL drivers for each kernel launch - on CloverLeaf 2D these kernels account for 5.4\% (nd\_range) and 8.7\% (flat) of the runtime, In contrast with MPI+OpenMP only 0.34\% of runtime is spent in boundary loops, and with OpenSYCL 2.5\% (nd\_range) and 1.24\% (flat) - this is because OpenSYCL maps to OpenMP at compile-time. For CloverLeaf 3D however, this flips around, with OpenSYCL spending up to 27\% of time in boundary loops. It is also noteworthy, that reductions take 6-7$\times$ more time with SYCL compared to OpenMP - here we had to use user-defined binary tree reductions as SYCL 2020's built-in reductions are not yet supported in OpenSYCL for this target, and had compilation issues with DPC++. The other applications show less variability across different parallelizations; the DPC++ compiler performs 10\% better on the more compute-intensive RTM and Acoustic applications compared to MPI/MPI+OpenMP due to better vectorization efficiency. The best performing implementations achieve between 42\% (RTM) and 77\% (CloverLeaf 2D) efficiency.

Moving on to the the AMD EPYC 9V33X (Genoa-X) processor, we were able to install and use the OneAPI C/C++ compilers to run SYCL code, however, as it is not optimized for this hardware, performance was not ideal. For the baseline, we used the AMD Optimizing C/C++ Compilers. For CloverLeaf 2D, both DPC++ (flat variant) and OpenSYCL (either variant) produced code that gave incorrect results. CloverLeaf 3D worked correctly, but the DPC++ compiler and OpenCL runtime gave significant overheads across the board - OpenSYCL with the nd\_range optimization was however able to slightly outperform both MPI and MPI+OpenMP. On the RTM application, MPI+OpenMP outperformed all other variants by 1.46-1.95$\times$, but for other codes the margins are reduced for various SYCL implementations. For the best variants, Genoa-X achieves up to 107\% efficiency on CloverLeaf 2D thanks to its large L3 cache, its lowest is 54\% on RTM - on average this platform gave the highest efficiency at 78\%.

Finally, on the Ampere Altra platform we were only able to compare OpenSYCL to MPI/OpenMP (the OneAPI toolkit only supports x86) - this architecture has a single NUMA node, so we didn't use MPI+OpenMP.  The SYCL implementations offer competitive performance here as well, being within 10-15\% of MPI or OpenMP for most applications except Acoustic, where auto-vectorization did not work for SYCL - but it did for MPI/OpenMP. OpenSBLI SN  failed to vectorize across all variants, achieving only 36\% efficiency. The less compute-intensive code achieved 75\% (CloverLeaf 2D), 56\% (CloverLeaf 3D) and 55\% (OpenSBLI SA). 

\subsection{Unstructured mesh application - MG-CFD}

\begin{figure*}[h]
    \centering
    \includegraphics[width=0.9\textwidth]{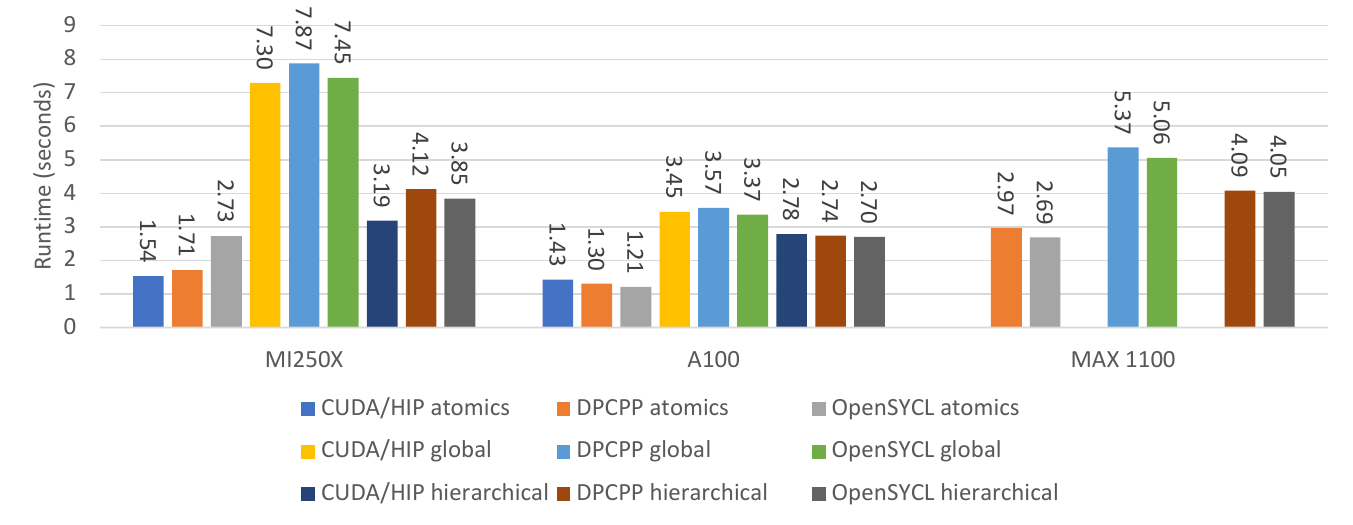} 
    \caption{Runtime of MG-CFD (Rotor37) on GPU architectures with different compilers and parallel implementations}
    \label{fig:mgcfd_gpu}
\end{figure*}

\begin{figure*}[h]
    \centering
    \includegraphics[width=0.9\textwidth]{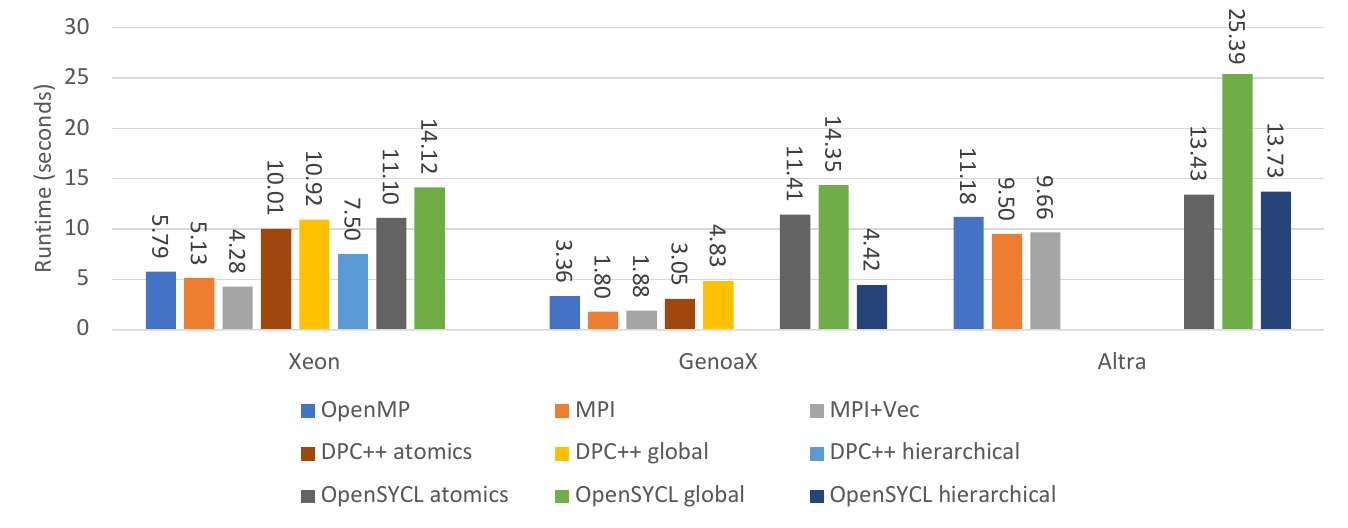} 
    \caption{Runtime of MG-CFD (Rotor37) on the CPU architectures with different compilers and parallel implementations}
    \label{fig:mgcfd_cpu}
\end{figure*}

Due to their complexity, unstructured mesh applications present more of a challenge to how they can be formulated in a programming model, then mapped to the hardware. As discussed, we have three distinct parallelization approaches that use different ways of avoiding race conditions - atomics, global coloring, and hierarchical coloring. GPUs have hardware support for fast floating-point atomics (though on the MI250X there are ``safe'' and ``unsafe'' ones - we used the unsafe ones where we could, and got correct results), CPUs only have support for generic atomics, and with a lower throughput, but they are functioning and give the correct results.

On Figure \ref{fig:mgcfd_gpu} we compare different parallelizations and compilers on GPU architectures - note that there is no ``native'' version to compare against on the Intel GPU Max 1100. Here, compilers work consistently, with a few notable differences. On the AMD MI250X with OpenSYCL, we could not access the unsafe atomics, therefore got significantly worse throughput with the safe atomics compared to HIP and DPC++. On the A100, SYCL implementations for all but one parallelizations and compilers outperformed native CUDA - with OpenSYCL+atomics 18\% faster than CUDA+atomics. Atomics throughput in the Max 1100 appears to be the limiting factor, as the performance of hierarchical and global coloring versions are in line with the performance differences to the A100.

Memory locality is the other key factor determining performance for MG-CFD. Given the good ordering of the mesh, the atomic version executes adjacent edges on adjacent threads/work items, giving good spatial locality for edge-based data and good temporal locality for vertex-based data. On the MI250X, the profiler reports this version reading 3500 bytes per wave (64 threads), and a 91\% hit rate in L2. For the global coloring approach adjacent edges have different colors, and therefore edges executed at the same time (which have the same color) will not be adjacent, and therefore both spatial and temporal locality will be poor. On the MI250X, this version reads 39000 bytes per wave, and achieves only 58\% hit rate in L2. The hierarchical coloring represents a middle ground: although different thread blocks/workgroups do not share data, threads/work items in the same workgroup do. On these GPUs the best performing block size was 256, on CPUs 4096. On the MI250X, we read 8600 bytes/wave and achieve an 83\% hit rate in L2. Similarly, the A100 GPU has L2 hit rates of 72\%/54\%/62\% respectively for the three schemes.

We calculate effective bandwidth in OP2 for each kernel as the total size of datasets accessed (multiplied by 2 if read-write), plus the size of mapping tables used, divided by execution time. Achieved effective bandwidth is then weighted averaged over different kernels. Once again, taking the best performing variants on each architecture, the A100 achieves 86\% efficiency, the MI250X 69\% efficiency, and the Max 1100 63\% efficiency.

The runtimes of MG-CFD on CPU architectures are shown in Figure \ref{fig:mgcfd_cpu}; here we see a very mixed picture. First, there are numerous SYCL variant and compiler combinations which failed to compile (with internal compiler errors, mostly OpenSYCL), crashed during execution, or produced incorrect results. These are the same variants as on the GPUs, where each worked and validated. Second, while DPC++ was able to vectorize the hierarchical formulation of SYCL, it was consistently slower than the non-vectorized version (which used a workgroup size of 1). In terms of execution scheme, this non-vectorizing hierarchical version corresponds to the MPI+OpenMP version, which also does not vectorize kernels with race conditions. Overall architectural efficiencies on the CPU platforms for the best implementations (auto-vectorizing MPI) are 108\% on the Xeon, 135\% on the Genoa-X, and 86\% on the Altra - these high values are due to data re-use across subsequent computational loops on the coarser levels of the multigrid, especially significant on the Genoa-X thanks to its large L3 cache ($2\times1.1$GB).

\subsection{Performance portability of SYCL}

Of all the programming approaches evaluated, we compared against non-portable baselines (MPI, MPI+OpenMP on CPUs, CUDA/HIP on GPUs). It is only SYCL code that one way or another was able to run on all platforms studied - one may argue that OpenMP is portable as well, given its support for offload, however different pragmas have to be used, and support in the open-source LLVM compilers for NVIDIA GPUs is still not great - we saw runtime errors for multiple applications (usually too many arguments trying to be passed at kernel launch). Technically, there was no single compiler and SYCL formulation that ran across all applications and architectures (with CloverLeaf 2D only working with DPC++ nd\_range on Genoa-X, and Altra not supporting DPC++), yet there is at least one compiler and SYCL formulation that works across all architectures and applications.

\begin{figure}[h]
    \centering
    \includegraphics[width=0.9\columnwidth]{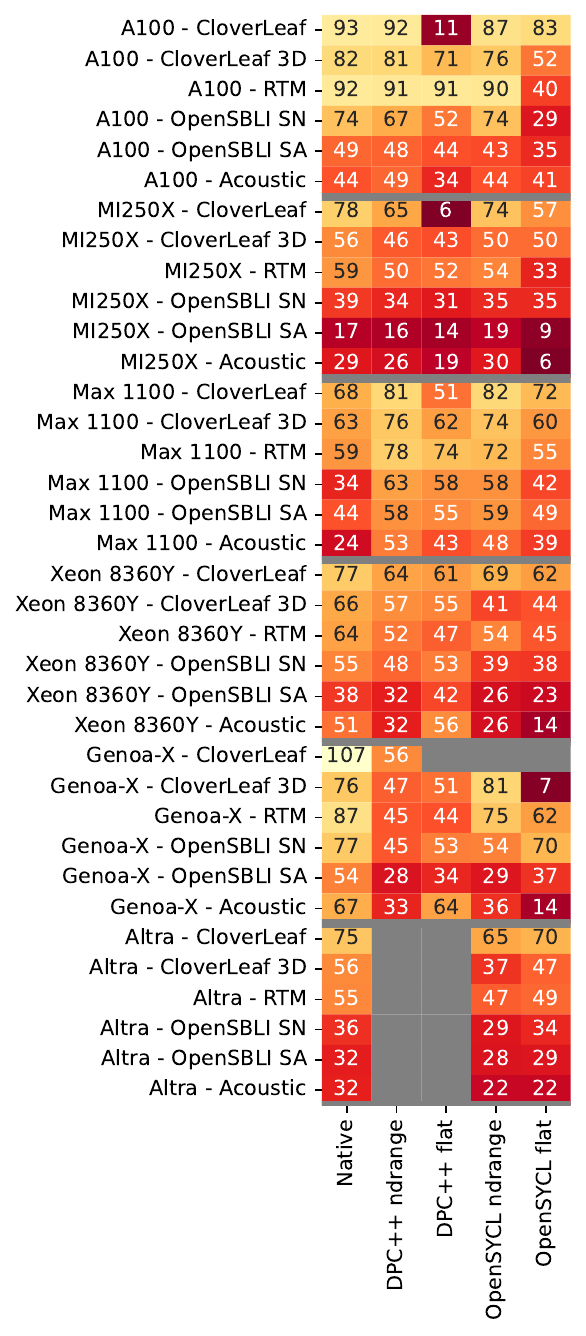} 
    \caption{Achieved architectural efficiency on structured mesh applications}
    \label{fig:heatmap_structured}
\end{figure}

\begin{figure}[h]
    \centering
    \includegraphics[width=0.9\columnwidth]{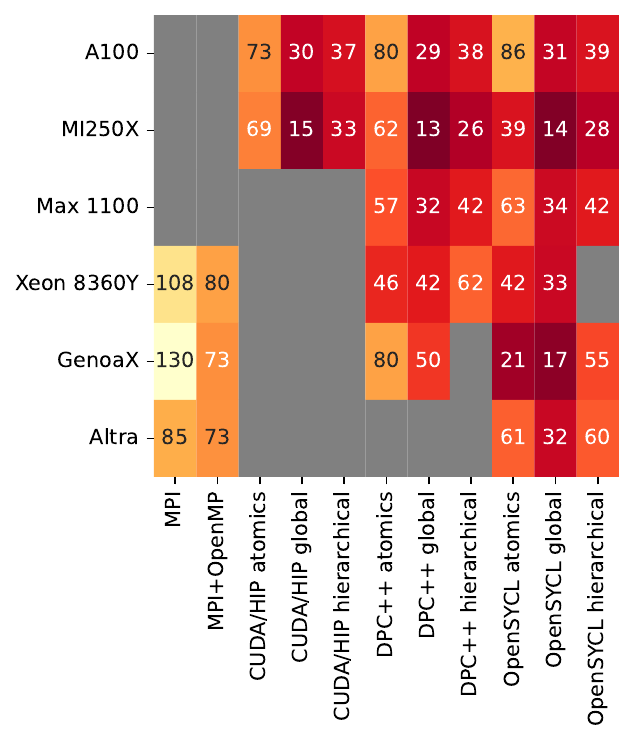} 
    \caption{Achieved architectural efficiency on the unstructured mesh application MG-CFD}
    \label{fig:heatmap_mgcfd}
\end{figure}

Figure \ref{fig:heatmap_structured} shows the architectural efficiency (percentage of peak memory bandwidth) for different combinations on structured mesh applications (note that ``native'' on the Intel Data Center GPU Max 1100 is OpenMP offload), and Figure \ref{fig:heatmap_mgcfd} for the unstructured mesh application MG-CFD. 

On structured meshes, the various native approaches on average achieve 59\% efficiency (with a standard deviation of 21\%) - in comparison DPC++ nd\_range achieves 54\% (std. 19\%) and OpenSYCL nd\_range achieves 52\% (std 21\%), which is very competitive for a single code variant across all these applications and platforms. With the flat variant this reduces to 47\% (std 19\%) and 41\% (std 19\%) with DPC++ and OpenSYCL respectively. There is a significant difference between CPU and GPU architectures however - nd\_range versions achieve 60\% efficiency on average across the different GPU architectures, whereas only 45\%-48\% on CPU architectures; there is only a small difference for the ``native'' approaches (56\% and 61\% respectively). The performance portability metric \cite{pennycook}, ignoring failing/unavailable variants gives 0.49 for DPC++ nd\_range and 0.46 for OpenSYCL nd\_range, which is reduced to 0.35 and 0.29 for the flat variants with DPC++ and OpenSYCL respectively.
Overall, SYCL delivers a very good trade-off in terms of portability and performance for structured mesh applications - part of which is its support for the ``flat'' formulation, which lets the runtime libraries pick the right workgroup shape for a given kernel and the target platform. While in a few cases this currently leads to significant performance degradation, which will likely be addressed in future releases, in most cases flat is shown to be competitive with a tuned nd\_range version. Furthermore, with an iterative development approach one can initially rely on the flat formulation, and move to nd\_range for the most critical kernels, optimizing the shape for the various targets.

The unstructured mesh application MG-CFD paints a more mixed picture. While on GPUs, SYCL performance is well in line with ``native'' approaches (43\% efficiency on average with native, 42\% with various SYCL versions), performance on CPUs is less consistent - even if we ignore the numerous aforementioned compiler/runtime issues. We can observe that performance on the Xeon 8360Y Processor with SYCL is not competitive (SYCL being 30\% slower than OpenMP). On GenoaX, the DPC++ atomics version (which does vectorize) is competitive with OpenMP, but not with pure MPI. OpenSYCL on Intel and AMD CPUs is also not competitive - its best is on GenoaX, where SYCL hierarchical is 31\% slower than OpenMP. On the Ampere Altra the matching hierarchical is still 21\% slower than OpenMP. The performance portability metric for OpenSYCL + atomics (which worked on all platforms) is 0.42, but when picking the best available compiler and SYCL variant combination on each platform, it improves to 0.67.

There is another important implication of these results; a single algorithmic variant/implementation may be portable across different hardware platforms, but aside from simpler cases, it will not be performance portable. This is well documented in the literature: massively parallel GPU architectures often require different algorithmic approaches than classical CPU architectures. SYCL does not solve this problem, but it does give its users a single programming model and environment to express these different variants and to target different architectures.

\section{Conclusions}

In this paper we have taken a thorough look at the performance portability of the SYCL programming model across most available modern HPC architectures, including Intel, AMD, and NVIDIA GPUs, as well as Intel, AMD, and ARM CPUs. On a set of proxy and production applications that are mainly bound by the performance of the memory system, we have contrasted different formulations of SYCL code, compiled with the two available compilers - DPC++ and OpenSYCL/hipSYCL. We have also compared these to established baseline implementations - MPI/MPI+OpenMP on CPUs and native CUDA/HIP on GPUs.

The results demonstrate very good portability for both structured and unstructured mesh computations on GPUs, with highly competitive performance compared to the native (non-portable) implementations. Across all applications and all platforms, the best native versions achieve on average 62.7\% architectural efficiency, and the best SYCL implementations achieve 59.1\%.

Performance portability is excellent on GPU architectures in particular, with native approaches only at 57.6\% efficiency and the best SYCL implementation at 62.7\%. In contrast, stability and performance on CPUs is still more of an issue, with several versions failing to compile, crashing, or giving incorrect results. Achieved efficiency with native approaches is on average at 67.8\% on CPUs, but with SYCL only at 55.5\%. SYCL implementations outperform native ones in a handful of notable cases - on GPUs (NVIDIA in particular), this is mainly due to the difference in the compiler stack, with LLVM applying more powerful optimizations, while on CPUs SYCL variants sometimes yielded more efficient vectorized implementations.

Overall utilizing SYCL is a good approach for performance portability if one needs this level of abstraction, with strong compiler support when targeting GPUs, but support for CPUs needs to be improved if SYCL is to become a truly versatile approach. While SYCL does not solve the performance portability challenge, it does provide a unified programming model capable of targeting most modern HPC architectures. Community support for SYCL may still prove to be a challenge, as currently only two compilers are broadly available (one by Intel, one an academic research project), which represents a risk in the longer term.

\begin{acks}
We are grateful for the support of the OneAPI Innovator program, and the advice and assistance of Mark Lubin, Xiao Zhu, and Rob Muller-Albrecht at Intel in particular. 

This research was supported by Rolls-Royce plc., and by the UK EPSRC (EP/S005072/1 – Strategic Partnership in Computational Science for Advanced Simulation and Modelling of Engineering Systems – ASiMoV). This work was also supported in part by the Hungarian Academy of Sciences under Grant POST-COVID2021-64. 

\end{acks}

%%
%% The next two lines define the bibliography style to be used, and
%% the bibliography file.
\bibliographystyle{ACM-Reference-Format}
\bibliography{sample-base}

\end{document}